\documentclass{article}
\usepackage{piner}
\usepackage[figuresright]{rotating}
\begin{document}

\submitted{Accepted by ApJ, Jan 16, 2003}
\title{The Speed and Orientation of the Parsec-Scale Jet in 3C\,279}

\author{B. Glenn Piner\altaffilmark{1}, Stephen C. Unwin\altaffilmark{2}, 
Ann E. Wehrle\altaffilmark{3}, Alma C. Zook\altaffilmark{4}, C. Megan Urry\altaffilmark{5},
\& Diane M. Gilmore\altaffilmark{6}}

\altaffiltext{1}{Department of Physics and Astronomy, Whittier College,
13406 E. Philadelphia Street, Whittier, CA 90608; gpiner@whittier.edu}

\altaffiltext{2}{Jet Propulsion Laboratory, Mail Code 301-486, 4800 
Oak Grove Drive, Pasadena, CA 91109; Stephen.C.Unwin@jpl.nasa.gov}

\altaffiltext{3}{Interferometry Science Center,
California Institute of Technology, Mail Code 301-486, 4800 Oak
Grove Drive, Pasadena, CA 91109; Ann.E.Wehrle@jpl.nasa.gov}

\altaffiltext{4}{Department of Physics and Astronomy, Pomona
College, Claremont, CA 91711; azook@pomona.edu}

\altaffiltext{5}{Yale University, Physics Department, P.O. Box 208121, New Haven, CT 06520;
meg.urry@yale.edu}

\altaffiltext{6}{Space Telescope Science Institute, Science Comp. \& Res. Support Div., 3700 San Martin Dr., 
Baltimore, MD 21218; dgilmore@stsci.edu}

\begin{abstract}
A high degree of relativistic beaming is inferred for the jets of blazars based
on several lines of evidence, but the intrinsic speed and angle of the jet to the line-of-sight 
for individual sources are difficult to measure.
We have calculated inverse-Compton Doppler factors for 3C~279 using the collection
of VLBI data (including high-resolution space VLBI data at low frequencies) 
recently published by us (as Wehrle et al. and Piner et al.), and the collection of multiwavelength
spectra recently published by Hartman et al.  From the Doppler factor and superluminal apparent speed,
we then calculate the Lorentz factor and angle to the line-of-sight of the parsec-scale relativistic jet.
We follow the method previously used by Unwin et al. for 3C~345 to model the jet components
as homogeneous spheres and the VLBI core as an unresolved inhomogeneous conical jet, using K\"{o}nigl's
formalism.  

The conical-jet model can be made to match both the observed 
X-ray emission and the VLBI properties of the core with a suitable choice of Doppler factor,
implying the core makes a significant contribution to the X-ray emission, in contrast to the situation for 3C~345,
where the jet components dominated the X-ray emission.
The parameters of the K\"{o}nigl models indicate the jet is particle dominated at the radii
that produce significant emission
(from $\sim$ 5 to 20 pc from the apex of the jet for most models),
and is not in equipartition.  
At the inner radius of the K\"{o}nigl jet the magnetic field is of order 0.1 G
and the relativistic-particle number density is of order 10 cm$^{-3}$.
The kinetic energy flux in the jet is of order
10$^{46} (1+k)$ ergs sec$^{-1}$, where $k$ is the ratio of proton to electron energy,
which implies a mass accretion rate of order $0.1 (1+k)/\eta M_{\odot}$ yr$^{-1}$,
where $\eta$ is the efficiency of conversion of mass to kinetic energy.

When all components are included in the calculation, then
on average the core produces about half of the
X-rays, with the other half being split between the long-lived component C4 and the brightest inner-jet component.
We calculate an average speed and angle to the line-of-sight for the region of the jet interior to 1 mas of
$v=0.992c$ ($\gamma=8$) and $\theta=4\arcdeg$, 
and an average speed and angle to the line-of-sight for C4 (at $r\approx3$ mas) of
$v=0.997c$ ($\gamma=13$) and $\theta=2\arcdeg$.
These values imply average Doppler factors of $\delta=12$ for the inner jet, and $\delta=21$ for C4.
\end{abstract}


\section{Introduction}
\label{intro}
The quasar 3C~279 ($z$=0.536) has been one of the most intensively studied quasars for several reasons.
It was the first radio source observed to exhibit the phenomenon of
apparent superluminal motion (Knight et al. 1971; Whitney et al. 1971; Cohen et al. 1971),
prompting continued study with VLBI through the 1970's and 1980's
(Cotton et al. 1979; Unwin et al. 1989; Carrara et al. 1993) and 1990's (Wehrle et al. 2001; hereafter Paper I).
A bright $\gamma$-ray flare was observed from 3C~279  in 1991 by the EGRET instrument shortly after the launch
of the {\em Compton Gamma Ray Observatory} (Hartman et al. 1992), 
leading to a great deal of multiwavelength coverage during
the following nine years.  3C~279 was one of the brightest $\gamma$-ray blazars detected during
the lifetime of the EGRET instrument (Hartman et al. 1999).  It is one of the prototypes for the class of
luminous ``red'' blazars (Sambruna 2000).

These studies have produced a large amount of data on 3C~279.
A number of nearly simultaneous multiwavelength spectra are available that show 3C~279
at various levels of activity.  Data previously published by
Maraschi et al. (1994), Hartman et al. (1996), and Wehrle et al. (1998) are compiled by
Hartman et al. (2001a) (hereafter H01).
Strong variability on timescales of a day or less has been observed in optical through
$\gamma$-ray bands (e.g., Wehrle et al. 1998; Hartman et al. 2001b).
In the radio regime, the variability timescale is longer, and flux density
monitoring at 4.8, 8.4 and 14.5 GHz, complete with polarization data,
has been obtained at the University of Michigan Radio Observatory
(Aller et al. 1985).  Monitoring at 22 and 37 GHz has been done at Mets\"{a}hovi
Observatory (Ter\"{a}sranta et al. 1992, 1998).
We have recently published a compendium of six years of VLBI images of 3C~279 at 22
and 43 GHz from 1991 to 1997 (Paper I), showing the kinematics of the parsec-scale jet.
In this paper we combine the VLBI data from Paper I with multiwavelength spectral information
to calculate the Doppler factor, orientation, and speed of 3C~279's parsec-scale jet.

The two-humped overall spectral energy distribution of blazars is most naturally
explained with a jet of relativistic electrons emitting a 
combination of synchrotron radiation for the radio
through optical-uv region, and inverse-Compton emission at higher
energies (e.g., Wehrle 1999).
This interpretation of the spectrum leads to derivation of physical conditions in the jet that 
include bulk relativistic motion.
Evidence for bulk relativistic motion in blazars comes from many sources, including the observed apparent
superluminal motions in blazar jets (Vermeulen \& Cohen 1994),
the transparency of blazar cores to high-energy $\gamma$-rays (Dondi \& Ghisellini 1995),
rapid flux variability (L\"{a}hteenm\"{a}ki \& Valtaoja 1999),
high VLBI core brightness temperatures (Tingay et al. 2001),
an excess of predicted over observed inverse-Compton emission (Ghisellini et al. 1993),
and arguments invoking equipartition or minimum energy requirements (Readhead 1994).

This bulk relativistic motion can be quantified by the ratio of observed to emitted frequency, or Doppler factor,
\begin{equation}
\delta =\frac{1}{\gamma (1-\beta \cos \theta)},
\end{equation}
where $\theta$ is the angle to the line-of-sight, $\beta =v/c$,
and $\gamma=(1-\beta^2)^{-1/2}$ is the bulk Lorentz factor. 
Although evidence for high values of $\delta$ is strong, calculating $\delta$ for any given
source is difficult, as we explain below.  Knowledge of $\delta$ is desirable because, together with the apparent superluminal
speed, it constrains both the bulk Lorentz factor (important for studying jet energetics) and the
angle of the jet to the line-of-sight (important for unification studies).
One method of calculating the Doppler factor is to use source properties measured from
VLBI images and multiwavelength spectra to predict the X-ray flux density that should be emitted
by the Synchrotron Self-Compton (SSC) process.  This calculation typically over-predicts the X-ray
flux density from the source.  By assuming the source is relativistically beamed with a certain $\delta$
(see equation~(\ref{spheredelta})),
the conflict between the predicted and the observed X-ray flux density can be eliminated (e.g., Marscher 1987; Ghisellini et al. 1993).  
If all of the X-rays from the source are not due to the SSC process from the component being considered,
then a lower limit to $\delta$ is obtained rather than a firm value. 

In practice, this method has many problems.  It depends on the assumed geometry of the emitting region,
and it depends sensitively on parameters that must be measured from multi-frequency VLBI images.
Many authors assume a homogeneous sphere geometry for the VLBI core
(e.g., Ghisellini et al. 1993; Mantovani et al. 2000).
In reality, a homogeneous sphere is a poor approximation to the VLBI core, because
it predicts a sharply-peaked synchrotron spectrum, and not the
flat spectrum over several decades in frequency shown by many sources.
A more realistic model for the VLBI core is an inhomogeneous conical jet, such as the model
by Blandford \& K\"{o}nigl (1979) and K\"{o}nigl (1981), which can reproduce
the observed multiwavelength spectral indices.  Estimates for parameters such as the synchrotron
turnover frequency of VLBI components also bring uncertainty into this calculation, and
L\"{a}hteenm\"{a}ki, Valtaoja, \& Wiik (1999) show that plotting SSC Doppler factors derived
by various authors from different data sets for the same sources results in an almost
pure scatter diagram, most likely due to varying assumptions about input parameters.

SSC Doppler factors have been calculated in detail perhaps only for 3C~345 by
Unwin et al. (1994, 1997).
They modeled 3C~345 as a superposition of an inhomogeneous conical jet 
(using K\"onigl's 1981 formalism) for
the VLBI core, and a series of homogeneous spheres for the VLBI components or ``blobs''.
They then used multi-frequency VLBI data, quasi-simultaneous X-ray data,
and the multiwavelength spectrum to constrain the Doppler factor of 3C~345.
In this paper, we apply the procedure used by Unwin et al. (1994, 1997) for
obtaining inverse-Compton Doppler factors to 3C~279.

Note that our goal in this paper is not to model the entire multiwavelength spectrum
and short-term variability of 3C~279 using the K\"{o}nigl model.  
We use the K\"{o}nigl model as a geometry somewhat more sophisticated 
than a homogeneous sphere to calculate what we hope will be an accurate measurement
of the SSC Doppler factor, in order to constrain the jet orientation and speed.
In particular, we make no effort to model the $\gamma$-ray portion of the spectrum,
because models including other radiation mechanisms, such as that of H01 and Ballo et al. (2002),
show that this is quite likely due to external Compton scattering and not SSC.
The model of H01 does indicate that the lower energy portion of
the inverse-Compton spectrum (the X-rays) is likely to be dominated by SSC emission.
We also consider only the comparatively ``quiescent'' emission from 3C~279, because the rapid variability
during flares may originate in components much smaller than the K\"{o}nigl
jet, presumably small blobs (like those in the H01 model) 
that are moving out through the inhomogeneous jet and are superposed
with the VLBI core in the VLBI images (see Paper I).  Note that even sophisticated homogeneous models
such as that of H01 do not come close to fitting the radio emission, showing that
some form of inhomogeneous jet component is required.

While X-ray emission on larger scales in blazar jets may be caused by inverse-Compton scattering
of the microwave background (e.g., Celotti, Ghisellini, \& Chiaberge 2001),
on the parsec scales considered in this paper the
synchrotron photon energy density is orders of magnitude higher than the microwave
background energy density (about $10^{-4}$ ergs cm$^{-3}$ compared to $10^{-10}$ ergs cm$^{-3}$,
using expressions from Celotti et al. 2001).
Because of this large difference in energy densities, inverse-Compton scattering
of the microwave background
is not considered further in this paper.

\section{Observations}
\label{obs}
The VLBI data used for this paper are taken from the 22 and 43 GHz VLBI observations
of 3C~279 presented in Paper I.  These data included 18 epochs spanning the time range from 1991
to 1997 (at 22 GHz) and 1995 to 1997 (at 43 GHz).  Earlier epochs in this sequence used
the Global VLBI Network and Mark II recording, later epochs used the NRAO VLBA telescopes
and correlator.  The reader is referred to Paper I for further discussion of the VLBI 
observations, and the VLBI images, model fits, and component identifications.
Two mosaics at 22 and 43 GHz from that paper are shown in Figures 1 and 2 for ease in identifying features.
Throughout the rest of this paper, C4 refers to the moving component approximately three milliarcseconds
from the core, C5 refers to the stationary component at one milliarcsecond from the core, and
C5a, C6, C7, C7a, C8, and C9 refer to the moving components that sequentially emerge from the
core during the course of the monitoring.  See Paper I for more details.
High resolution VLBI data at low frequencies is needed
to constrain the optically thick VLBI component spectral indices, so we also make use of the
1.6 and 5 GHz space VLBI (VSOP) observations of 3C~279 from Piner et al. (2000).

\begin{figure*}
\plotfiddle{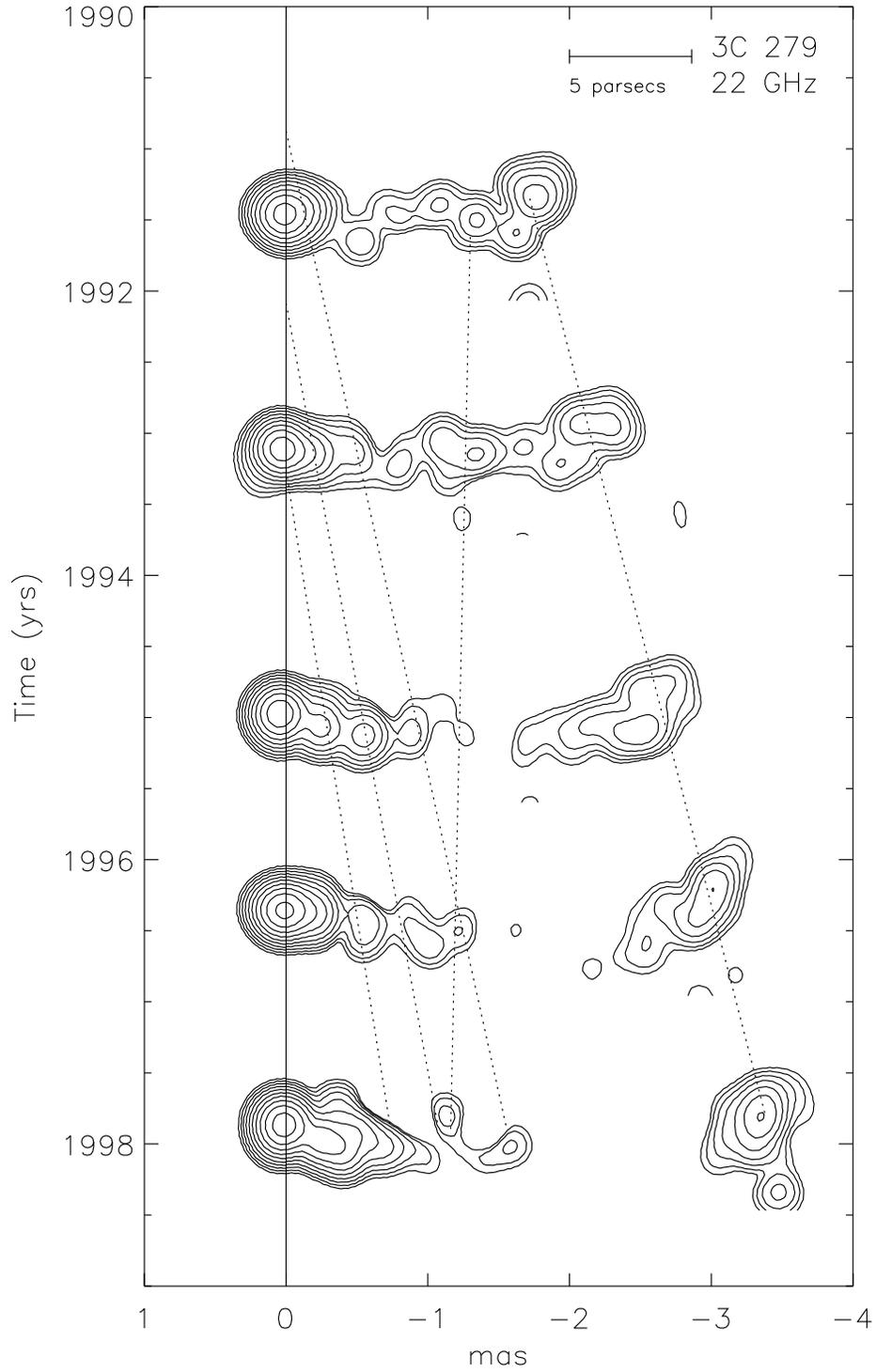}{8.0 in}{0}{75}{75}{-232}{0}
\caption{Time-series mosaic of a selection of 22 GHz VLBI images of 3C\,279.  Epochs
1991 Jun 24, 1993 Feb 17, 1994 Sep 21, 1996 May 13, and 1997 Nov 16 are shown.
The images have been restored with a circular 0.2 mas beam without residuals and
rotated 25$\arcdeg$ counterclockwise.  The lowest contour is 25 mJy beam$^{-1}$;
subsequent contours are a factor of two higher than the previous contour.
The solid line indicates the position of the presumed stationary core.  The dotted lines
represent the best fits to the model-fit Gaussian positions vs. time.
From right to left these lines represent C4, C5, C5a, C6, and C7.  Some lines have been extended
before and after model-fit detections to show speculative zero-separation epochs and
later positions.}
\end{figure*}

\begin{figure*}
\plotfiddle{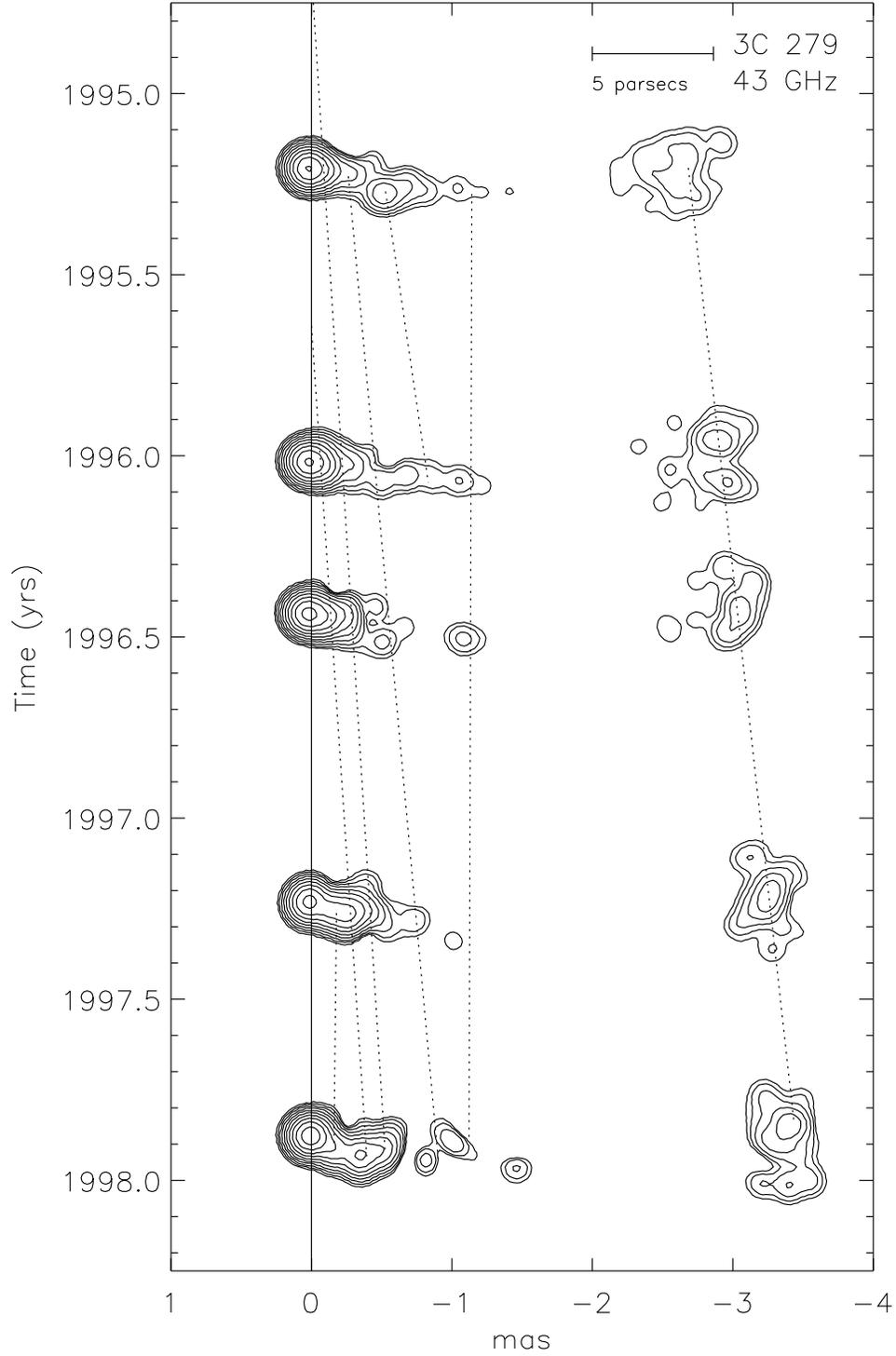}{8.0 in}{0}{75}{75}{-232}{0}
\caption{Time-series mosaic of a selection of 43 GHz VLBI images of 3C\,279.  Epochs
1995 Mar 19, 1996 Jan 7, 1996 Jun 9, 1997 Mar 29, and 1997 Nov 16 are shown.
The images have been restored with a circular 0.15 mas beam without residuals and
rotated 25$\arcdeg$ counterclockwise.  The lowest contour is 25 mJy beam$^{-1}$;
subsequent contours are a factor of two higher than the previous contour.
The solid line indicates the position of the presumed stationary core.  The dotted lines
represent the best fits to the model-fit Gaussian positions vs. time.
From right to left these lines represent C4, C5, C6, C7, C7a, C8, and C9.  Some lines have been extended
before model-fit detections to show speculative zero-separation epochs.}
\end{figure*}

\begin{sidewaystable*}
\caption{Spectral Indices and Breakpoints Derived from Multiwavelength Campaign Data, and
Reference X-Ray Measurements}
\label{mwspectra}
\begin{center}
{\footnotesize \begin{tabular}{l l r r r r r r r r r r} \tableline \tableline \\[-5pt]
& & & & & & \multicolumn{1}{c}{$\nu_{t}^{\ast}$} & \multicolumn{1}{c}{$S_{t}^{\ast}$} & \multicolumn{1}{c}{$\nu_{b}^{\diamond}$}
& \multicolumn{1}{c}{$S_{b}^{\diamond}$} & \multicolumn{1}{c}{$\nu_{x}^{\circ}$} & \multicolumn{1}{c}{$S_{x}^{\circ}$} \\
\multicolumn{1}{c}{Date Range} & \multicolumn{1}{c}{Epoch$^{\dagger}$} & \multicolumn{1}{c}{$\alpha_{s1}^{\ddagger}$}
& \multicolumn{1}{c}{$\alpha_{s2}^{\ddagger}$} & \multicolumn{1}{c}{$\alpha_{s3}^{\ddagger}$} & \multicolumn{1}{c}{$\alpha_{c2}^{\ddagger}$}
& \multicolumn{1}{c}{(Hz)} & \multicolumn{1}{c}{(Jy)} & \multicolumn{1}{c}{(Hz)} & \multicolumn{1}{c}{(Jy)}
& \multicolumn{1}{c}{(keV)} & \multicolumn{1}{c}{($\mu$Jy)} \\ \tableline \\
1991 Jun 15 --- 1991 Jun 28 & P1  & $0.22\pm0.01$ & $-0.57\pm0.05$ & $-1.29\pm0.06$ & $-0.67\pm0.03$
& $6.8\pm^{0.5}_{0.5}\times10^{10}$ & $18.6\pm^{0.5}_{0.5}$ & $1.1\pm^{0.5}_{0.3}\times10^{13}$ & $1.0\pm^{0.5}_{0.3}$    & 10.0 & 0.90 \\[5pt]
1992 Dec 22 --- 1993 Jan 12 & P2  & $0.24\pm0.02$ & $-0.81\pm0.03$ & $-1.66\pm0.03$ & $-0.91\pm0.07$
& $6.5\pm^{0.3}_{0.3}\times10^{10}$ & $20.3\pm^{0.5}_{0.5}$ & $1.8\pm^{0.4}_{0.4}\times10^{13}$ & $0.21\pm^{0.08}_{0.05}$ & 1.0  & 1.00 \\[5pt]
1993 Dec 13 --- 1994 Jan 3  & P3b & $0.33\pm0.02$ & $-0.59\pm0.07$ & $-1.28\pm0.03$ & $-0.75\pm0.17$
& $5.9\pm^{0.5}_{0.4}\times10^{10}$ & $27.3\pm^{1.0}_{0.9}$ & $2.1\pm^{0.8}_{0.6}\times10^{12}$ & $3.4\pm^{1.3}_{0.9}$    & 2.3  & 0.62 \\[5pt]
1994 Nov 29 --- 1995 Jan 10 & P4  & $0.22\pm0.01$ & $-0.63\pm0.05$ & $-1.61\pm0.17$ & $-0.65\pm0.11$
& $4.2\pm^{0.2}_{0.2}\times10^{10}$ & $21.2\pm^{0.5}_{0.5}$ & $5.2\pm^{5.0}_{2.6}\times10^{12}$ & $1.0\pm^{0.8}_{0.5}$    & 2.4  & 0.49 \\[5pt]
1996 Jan 16 --- 1996 Jan 30 & P5a & $0.33\pm0.03$ & $-0.50\pm0.12$ & $-1.72\pm0.10$ & $-0.66\pm0.02$
& $5.7\pm^{1.4}_{1.1}\times10^{10}$ & $26.5\pm^{2.2}_{2.0}$ & $1.1\pm^{0.8}_{0.5}\times10^{13}$ & $1.9\pm^{1.8}_{0.9}$    & 10.0 & 0.33 \\[5pt]
1996 Dec 10 --- 1997 Jan 28 & P6a & $0.33\pm0.03$ & $-0.73\pm0.08$ & $-1.53\pm0.07$ & $-0.67\pm0.02$
& $4.2\pm^{0.3}_{0.3}\times10^{10}$ & $28.9\pm^{1.0}_{0.9}$ & $1.4\pm^{1.0}_{0.6}\times10^{13}$ & $0.43\pm^{0.44}_{0.22}$ & 9.5  & 0.23 \\[5pt]
1997 Jun 17 --- 1997 Jun 24 & P6b & $0.39\pm0.01$ & $-0.73\pm0.06$ & $-1.99\pm0.20$ & $-0.70\pm0.10$
& $4.3\pm^{0.2}_{0.2}\times10^{10}$ & $33.8\pm^{0.8}_{0.8}$ & $8.3\pm^{4.4}_{2.9}\times10^{13}$ & $0.14\pm^{0.11}_{0.07}$ & 3.0  & 0.48 \\[5pt]
\tableline\\
\end{tabular}}
\end{center}
{\footnotesize $\dagger$ Epoch identification in H01.}\\[5pt]
{\footnotesize $\ddagger$ $\alpha_{s1}$ is the optically thick synchrotron spectral index
below the turnover frequency, $\alpha_{s2}$ is the synchrotron spectral index above
the turnover frequency but below the break frequency,\\ $\alpha_{s3}$ is the synchrotron
spectral index above the synchrotron break frequency, and $\alpha_{c2}$ is the inverse-Compton
spectral index in the X-ray region of the spectrum.}\\[5pt]
{\footnotesize $\ast$ $\nu_{t}$ and $S_{t}$ are the fitted synchrotron turnover frequency and flux density at the turnover frequency.}\\[5pt]
{\footnotesize $\diamond$ $\nu_{b}$ and $S_{b}$ are the fitted synchrotron break frequency and flux density at the break frequency.}\\[5pt]
{\footnotesize $\circ$ $\nu_{x}$ and $S_{x}$ are the frequency and flux density of the reference X-ray measurement.}
\end{sidewaystable*}

\begin{figure*}[!t]
\plotone{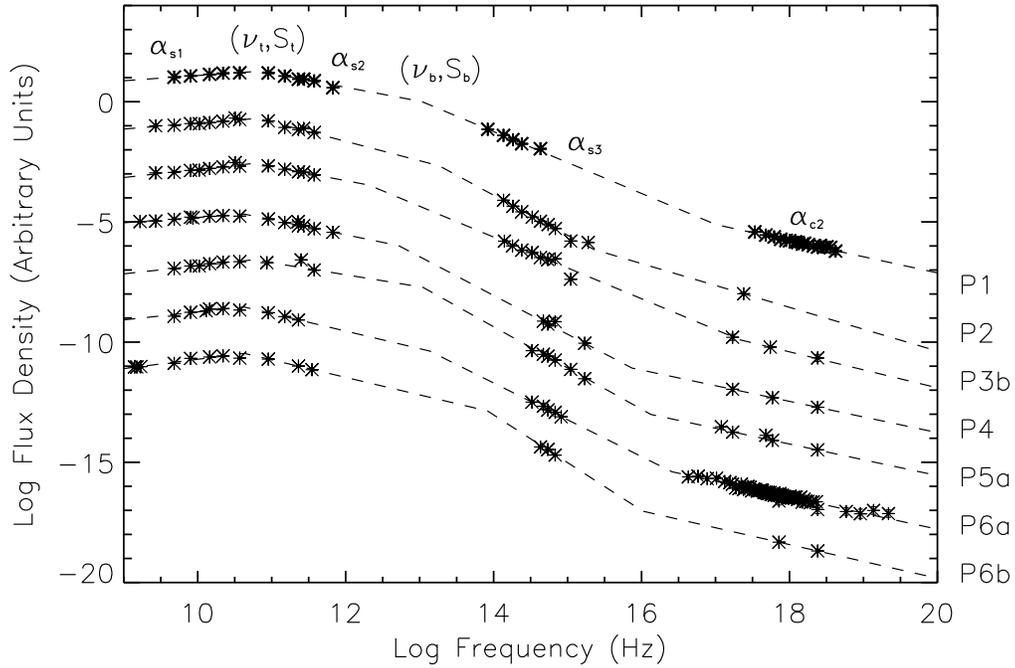}
\caption{Fits of the multiwavelength spectra from Table~\ref{mwspectra} with four power laws.
The flux density scale in Janskys is correct for the top (P1) spectrum, after that each successive spectrum
is offset by two orders of magnitude in flux density.
The error bars are smaller than the plotting symbols.
The parameters of the fits are given in Table~\ref{mwspectra}, and the
epoch designations from Table~\ref{mwspectra} are listed to the right of the figure.
The spectral indices and breakpoints are labeled for the P1 spectrum.}
\end{figure*}

Multiwavelength spectra of 3C~279 are needed to constrain inputs to both the
homogeneous sphere and inhomogeneous jet models.  For this purpose, we use the set of
contemporaneous multiwavelength spectra compiled by H01.  Eleven such spectra spanning the
years 1991 to 2000 are presented in that paper.  We fit
these spectra using four broken power laws applied to 
the different regions of the spectrum, to provide the appropriate inputs to the models.
Table~\ref{mwspectra} lists our power-law fits to the multiwavelength spectra of H01, for epochs from
1991 to 1997.  We exclude H01's epochs P3a (which has no X-ray data) and P5b (the very
large flare from 1996 February).  Spectral indices follow the naming convention
of K\"{o}nigl (1981): $\alpha_{s1}$ is the optically thick synchrotron spectral index
below the turnover frequency, $\alpha_{s2}$ is the synchrotron spectral index above 
the turnover frequency but below the break frequency, $\alpha_{s3}$ is the synchrotron
spectral index above the synchrotron break frequency, and $\alpha_{c2}$ is the inverse-Compton
spectral index in the X-ray region of the spectrum.  
The sign convention for spectral indices in this paper is $S\propto\nu^{+\alpha}$.
Errors on the fitted spectral indices were calculated from the errors in the measured flux densities
given in H01.  When only one X-ray measurement was available at a given epoch in H01, the
X-ray spectral index was acquired from the original X-ray paper referenced by H01.
Table~\ref{mwspectra} also gives the fitted turnover frequency and flux density at the turnover frequency
($\nu_{t}$ and $S_{t}$), the fitted break frequency and flux density at the break frequency
($\nu_{b}$ and $S_{b}$), and a reference X-ray measurement specified by $\nu_{x}$ and $S_{x}$
that we attempt to match with the models.
Figure 3 shows the power-law fits to the spectra from Table~\ref{mwspectra}.

\section{Calculation of SSC Doppler Factors}
The VLBI morphology and multiwavelength data cannot be adequately explained by
either an inhomogeneous jet ($\S$~\ref{inhomjet}) or a homogeneous sphere
(or spheres) model ($\S$~\ref{sphere}) alone.  In a combined model ($\S$~\ref{combined}),
we require the Doppler factor to be the same for the jet and spheres; the spheres
can therefore be regarded as approximations to dense clumps propagating along with the bulk jet.

\subsection{Homogeneous Sphere Model}
\label{sphere}
Possibly the two simplest blazar models are the homogeneous sphere model (Gould 1979)
and the inhomogeneous jet (Blandford and K\"{o}nigl 1979; K\"{o}nigl 1981).
In both models, the source is assumed to be moving relativistically at a small
angle to the line-of-sight, and the electrons responsible for the synchrotron
emission have a power-law distribution in energy.  As implied by their names, the
homogeneous sphere model assumes that the photon density within the spherical
source is uniform, while the inhomogeneous jet model assumes that the source is a
conical jet in which the magnetic field and electron number density have power-law
dependences on the distance from the apex of the jet.

The homogeneous sphere model, the simpler of the two, requires six observables to
determine the Doppler factor and the angle between the source velocity and the line-of-sight.
Three of these characterize the synchrotron spectrum: the spectral index
of the optically thin spectrum, and the flux density and frequency at which the
sphere first becomes optically thick (the ``turnover'' flux density and frequency).
Adding the angular size should in principle
determine the entire spectrum of the source, if the X-rays are the result of
SSC scattering of the synchrotron photons.  In practice, such a
calculation predicts X-ray flux densities that are several orders of magnitude
larger than observed, if the bulk motion of the source is assumed to be
non-relativistic.  This ``Compton catastrophe'' can be avoided if the source has a
relativistic bulk velocity, and in this case the observed X-ray flux density (the
fifth observable) provides a constraint that determines the Doppler factor.
Together with the apparent superluminal speed (the sixth observable), this determines both the bulk velocity and
the angle to the line-of-sight.

\begin{figure*}[!t]
\plotone{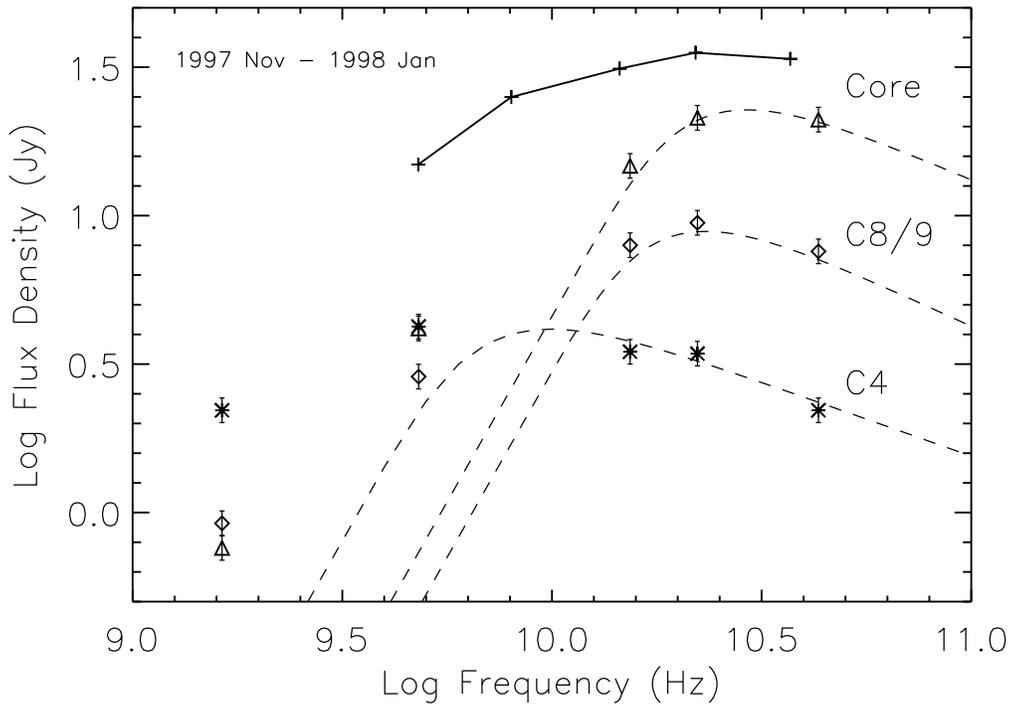}
\caption{Five-frequency VLBI spectra of 3C~279 core and jet
components from observations between 1997 November and 1998 January.  Three bright components
are shown: the VLBI core (triangles), the blended component C8/9 (diamonds),
and C4 (asterisks).  The single-dish spectrum from Michigan and
Mets\"{a}hovi monitoring is also shown (plus signs).  Three sample
homogeneous sphere spectra that match the three highest frequencies
for each component are also shown.}
\end{figure*}

Figure 4 shows sample homogeneous sphere spectra plotted with five-frequency
VLBI data from late 1997 and early 1998.  The data at 15, 22, and 43 GHz
are from the VLBA observation on 1997 Nov 16 (Paper I), while the data
at 1.6 and 5 GHz are from the VSOP observation on 1998 January 9 (Piner et al. 2000).
The three brightest components in the source at that epoch (the core,
the long-lived bright component C4, and the newly emerged components C8/9,
blended except at 43 GHz) are plotted on this figure.
Homogeneous sphere spectra are calculated from 
\begin{equation}
\label{spherespec}
S(\nu)=S_{0}(\nu_{t}/\nu_{0})^{\alpha}(\nu/\nu_{t})^{2.5}(1-e^{-\tau}),
\end{equation}
where $\nu_{t}$ is the turnover frequency, $\alpha$ is the optically
thin spectral index, $S_{0}$ is the flux density at frequency $\nu_{0}$,
and $\tau=(\nu_{t}/\nu)^{2.5-\alpha}$.  The sphere spectra
have been adjusted to go through the three highest frequency observations
for each component.  Inspection of Figure 4 shows
that homogeneous spheres cannot fit the spectra of either the core or the
jet components.  This is because a homogeneous sphere has an optically thick
spectral index of 2.5, while the actual components have optically thick indices
much less than this ($\sim 1.0$).  Evidently there are inhomogeneities in the ``components''
that broaden the self-absorption turnover and cause the spectral index below
the turnover to be less than 2.5.  The high-resolution space VLBI
data at low-frequencies was crucial to obtaining this result; ground-based images at low frequencies
blend the components so a spectral dissection like that in Figure 4 cannot be obtained.
The primary advantage of the homogeneous sphere model is its simplicity; it requires
a minimum number of observables to constrain the Doppler factor.
For this reason, we proceed with the homogeneous sphere calculation in this section,
with the expectation that corrections due to inhomogeneities (at least for the jet components)
will be small.  An alternative geometry
for the VLBI core (a conical jet) is tried in the next subsection.
An alternative geometry for jet components (oblique shocks) is considered by, e.g., Aller et al. (2001).

The Doppler factor of a homogeneous sphere that produces SSC X-rays of
flux density $S_{x}$ Jy at frequency $\nu_{x}$ (in keV) is (Ghisellini et al. 1993):
\begin{equation}
\delta=f(\alpha)S_{t}\left[\frac{\ln(\nu_{b}/\nu_{t})\nu_{x}^{\alpha}}
{S_{x}\xi_{d}^{6-4\alpha}\nu_{t}^{5-3\alpha}}\right]^{1/(4-2\alpha)}(1+z),
\label{spheredelta}
\end{equation}
where $f(\alpha)\approx-0.08\alpha+0.14$, $S_{t}$ is the flux density (in Jy) at the turnover
frequency $\nu_{t}$ (in GHz) obtained by extrapolating the straight-line optically thin slope (Marscher 1987), 
$\nu_{b}$ is the synchrotron break frequency in GHz, 
$\xi_{d}$ is the angular diameter of the VLBI component (in mas),
and $\alpha$ is the optically thin spectral index.
Equation~(\ref{spheredelta}) finds the Doppler factor for which the given spherical
component produces all the X-ray flux density from the source.  Since the other
components must also contribute X-rays at some level, the Doppler factor found
by this method is a lower limit for the given component.  To put this
method into practice to find an actual Doppler factor for 3C~279, 
we must assume the source has a constant Doppler factor along its jet.
We then apply equation~(\ref{spheredelta}) to find the lower limit to the Doppler factor
for each component at a given epoch.  The actual Doppler factor at that epoch 
must then be somewhat larger than the highest lower limit found from equation~(\ref{spheredelta}).
We then calculate the Doppler factor that will
produce the observed X-ray flux density from the sum of the X-rays from all components,
and what percentage of the observed X-ray flux density is produced by each component.

\begin{table*}[!t]
\caption{Lower Limits to the Doppler Factor Computed
Using the Homogeneous Sphere Model}
\label{spherecalc}
\begin{center}
\begin{tabular}{l l l r r r r r} \tableline \tableline
& & & \multicolumn{1}{c}{$S_{t}$} & \multicolumn{1}{c}{$\nu_{t}$} & & \multicolumn{1}{c}{$\xi_{d}$} & \\
\multicolumn{1}{c}{VLBI Epoch} & \multicolumn{1}{c}{H01 Epoch} & \multicolumn{1}{c}{Comp.} & \multicolumn{1}{c}{(Jy)} &
\multicolumn{1}{c}{(GHz)} & \multicolumn{1}{c}{$\alpha$} & \multicolumn{1}{c}{(mas)} & \multicolumn{1}{c}{$\delta$} \\ \tableline \\
1991 Jun 24 & P1 (1991 Jun 15 --- 1991 Jun 28)  & C4     & 4.1  & 7.0  & -0.5  & 0.25 & 15.7 \\
1992 Jun 14 & P2 (1992 Dec 22 --- 1993 Jan 12)  & Core   & 20.5 & 65.0 & -0.81 & 0.20 & 5.9  \\
            &                                   & C4     & 4.4  & 7.0  & -0.5  & 0.49 & 7.3  \\
1992 Nov 10 & P2 (1992 Dec 22 --- 1993 Jan 12)  & Core   & 18.6 & 65.0 & -0.81 & 0.14 & 10.0 \\
            &                                   & C6     & 4.4  & 22.2 & -0.6  & 0.25 & 4.2  \\
            &                                   & C4     & 6.2  & 7.0  & -0.5  & 0.83 & 4.4  \\
1993 Feb 17 & P2 (1992 Dec 22 --- 1993 Jan 12)  & Core   & 22.7 & 65.0 & -0.81 & 0.16 & 9.1  \\
            &                                   & C4     & 5.2  & 7.0  & -0.5  & 0.79 & 4.0  \\
1993 Nov 8  & P3b (1993 Dec 13 --- 1994 Jan 3)  & Core   & 24.7 & 58.7 & -0.59 & 0.16 & 11.8 \\
1994 Mar 2  & P3b (1993 Dec 13 --- 1994 Jan 3)  & Core   & 23.5 & 58.7 & -0.59 & 0.14 & 13.8 \\
            &                                   & C7     & 6.7  & 22.2 & -0.6  & 0.20 & 8.7  \\
1994 Jun 12 & P4 (1994 Nov 29 --- 1995 Jan 10)  & Core   & 28.0 & 42.4 & -0.63 & 0.18 & 18.7 \\
1994 Sep 21 & P4 (1994 Nov 29 --- 1995 Jan 10)  & Core   & 26.2 & 42.4 & -0.63 & 0.18 & 17.5 \\
1995 Jan 4  & P4 (1994 Nov 29 --- 1995 Jan 10)  & Core   & 21.8 & 42.4 & -0.63 & 0.13 & 25.9 \\
1995 Feb 25 & P4 (1994 Nov 29 --- 1995 Jan 10)  & Core   & 17.6 & 42.4 & -0.63 & 0.16 & 14.0 \\
            &                                   & C7     & 3.1  & 22.2 & -0.6  & 0.11 & 11.8 \\
1995 Mar 19 & P4 (1994 Nov 29 --- 1995 Jan 10)  & Core   & 22.7 & 42.4 & -0.63 & 0.14 & 21.8 \\
1996 Jan 7  & P5a (1996 Jan 16 --- 1996 Jan 30) & Core   & 23.2 & 56.6 & -0.5  & 0.11 & 26.1 \\
            &                                   & C4     & 4.2  & 7.0  & -0.5  & 0.40 & 9.6  \\
1996 May 13 & P5a (1996 Jan 16 --- 1996 Jan 30) & Core   & 24.3 & 56.6 & -0.5  & 0.11 & 27.2 \\
            &                                   & C7a/C8 & 8.7  & 22.2 & -0.6  & 0.20 & 11.4 \\
1996 Jun 9  & P5a (1996 Jan 16 --- 1996 Jan 30) & Core   & 25.3 & 56.6 & -0.5  & 0.07 & 54.4 \\
1997 Jan 15 & P6a (1996 Dec 10 --- 1997 Jan 28) & Core   & 26.2 & 42.4 & -0.73 & 0.12 & 30.1 \\
            &                                   & C8     & 14.5 & 22.2 & -0.6  & 0.22 & 17.9 \\
            &                                   & C4     & 6.6  & 7.0  & -0.5  & 0.31 & 24.6 \\
1997 Mar 29 & P6a (1996 Dec 10 --- 1997 Jan 28) & Core   & 20.7 & 42.4 & -0.73 & 0.10 & 31.0 \\
            &                                   & C8/9   & 16.1 & 22.2 & -0.6  & 0.18 & 26.8 \\
            &                                   & C4     & 4.6  & 7.0  & -0.5  & 0.20 & 34.5 \\
1997 Jul 16 & P6b (1997 Jun 17 --- 1997 Jun 24) & Core   & 24.8 & 42.8 & -0.73 & 0.13 & 26.8 \\
            &                                   & C8/9   & 16.9 & 22.2 & -0.6  & 0.31 & 12.4 \\
            &                                   & C4     & 7.8  & 7.0  & -0.5  & 0.43 & 17.0 \\
1997 Nov 16 & P6b (1997 Jun 17 --- 1997 Jun 24) & Core   & 41.4 & 42.8 & -0.73 & 0.14 & 38.1 \\
            &                                   & C4     & 8.9  & 7.0  & -0.5  & 0.40 & 22.3 \\ \tableline
\end{tabular}
\end{center}
\end{table*}

\begin{table*}[!t]
\caption{X-Ray Production in the Homogeneous Sphere Model}
\label{spheretot}
\begin{center}
\begin{tabular}{l r r r r r r r r} \tableline \tableline
& & \multicolumn{7}{c}{Fraction of Total X-Rays by Component} \\
Epoch & $\delta$ & Core & C4   & C6   & C7   & C7a/C8$^{\dagger}$ & C8   & C8/9$^{\dagger}$ \\ \tableline \\
1991 Jun 24 & 16 & ...  & 1.00 & ...  & ...  & ...    & ...  & ...  \\
1992 Jun 14 &  8 & 0.22 & 0.78 & ...  & ...  & ...    & ...  & ...  \\
1992 Nov 10 & 10 & 0.97 & 0.02 & 0.01 & ...  & ...    & ...  & ...  \\
1993 Feb 17 &  9 & 0.98 & 0.02 & ...  & ...  & ...    & ...  & ...  \\
1993 Nov 8  & 12 & 1.00 & ...  & ...  & ...  & ...    & ...  & ...  \\
1994 Mar 2  & 14 & 0.92 & ...  & ...  & 0.08 & ...    & ...  & ...  \\
1994 Jun 12 & 19 & 1.00 & ...  & ...  & ...  & ...    & ...  & ...  \\
1994 Sep 21 & 18 & 1.00 & ...  & ...  & ...  & ...    & ...  & ...  \\
1995 Jan 4  & 26 & 1.00 & ...  & ...  & ...  & ...    & ...  & ...  \\
1995 Feb 25 & 15 & 0.71 & ...  & ...  & 0.29 & ...    & ...  & ...  \\
1995 Mar 19 & 22 & 1.00 & ...  & ...  & ...  & ...    & ...  & ...  \\
1996 Jan 7  & 26 & 0.99 & 0.01 & ...  & ...  & ...    & ...  & ...  \\
1996 May 13 & 27 & 0.99 & ...  & ...  & ...  & 0.01   & ...  & ...  \\
1996 Jun 9  & 54 & 1.00 & ...  & ...  & ...  & ...    & ...  & ...  \\
1997 Jan 15 & 32 & 0.69 & 0.26 & ...  & ...  & ...    & 0.05 & ...  \\
1997 Mar 29 & 39 & 0.30 & 0.55 & ...  & ...  & ...    & ...  & 0.15 \\
1997 Jul 16 & 27 & 0.89 & 0.09 & ...  & ...  & ...    & ...  & 0.02 \\
1997 Nov 16 & 39 & 0.94 & 0.06 & ...  & ...  & ...    & ...  & ...  \\ [5pt] \tableline
\end{tabular}
\end{center}
\hspace{1.30in} $\dagger$ Refers to a blended component.
\end{table*}

In Table~\ref{spherecalc} we show the inputs to equation~(\ref{spheredelta}) and the
resulting Doppler factor lower limits at
each 22 GHz VLBI epoch, for those components found to produce a significant percentage ($>1\%$) 
of the X-ray flux density at that epoch.
The parsec-scale radio structure of 3C~279 consists of the compact core, the bright long-lived
component C4 located about 3 mas from the core at these epochs, and a series of short-lived inner-jet
components that have all faded by the time
they reached about 1 mas from the core.  The inner-jet components have similar spectra, typified by the
spectra of the blended component C8/9 in Figure 4.  We use C8/9 as a guide for all of the inner-jet
components, and take them to have $\nu_{t} = 22$ GHz and $\alpha = -0.6$ (see Figure 4).
For component C4, we take $\nu_{t} = 7$ GHz and $\alpha=-0.5$, based on power-law fits to the
optically thick and thin portions of the spectrum in Figure 4.  For the core turnover frequency
and spectral index, we use the turnover frequency
and spectral index ($\alpha_{s2}$) of the multiwavelength spectrum from the H01 epoch closest
in time to the VLBI epoch (see Table~\ref{mwspectra}), since the core dominates the high-radio-frequency spectrum.
$S_{t}$ for the jet components is 
calculated from the observed flux density at the turnover frequency (for the inner-jet components), or
extrapolated to 7 GHz (for C4), with the opacity correction described by Marscher (1987) applied
($S_{t}=S_{obs}e^{\tau_{m}}$), where $\tau_{m}$ is tabulated by Marscher (1987).
The core $S_{t}$ is taken to be the greater of $S_{t}$ from the closest multiwavelength
spectrum with extrapolations of jet component flux densities subtracted, or the observed 22 GHz (or 43 GHz if available) VLBI
core flux density with the opacity correction applied.  $\xi_{d}$ is taken to be the VLBI model-fit size (or size upper
limit if the fitted size is zero), multiplied by 1.8 to convert from the
model fit Gaussian FWHM to the diameter of an optically thin sphere (Pearson 1995).
$\nu_{b}$, $\nu_{x}$, and $S_{x}$ are taken from Table~\ref{mwspectra}, from the multiwavelength spectrum closest in
time to the VLBI epoch.
Note equation~(\ref{spheredelta}) is very insensitive to these three values (e.g., $\delta \propto S_{x}^{0.2}$ 
for $\alpha=-0.5$), X-ray variability between
the X-ray and VLBI observation should not have a large affect
(e.g., if the X-ray flux density increases by a factor of 10, the calculated Doppler
factor increases by a factor of 1.6).

In Table~\ref{spheretot} we calculate the Doppler factor for each epoch, assuming
that all components at a given epoch have the same Doppler factor, and that the total X-ray
emission from all components must equal the observed X-ray emission.
This Doppler factor will 
be at least slightly higher than the highest lower limit found 
for that epoch in Table~\ref{spherecalc}, since each lower limit in Table~\ref{spherecalc}
assumed the component under consideration was the sole source of the X-ray emission.
Errors in the turnover frequency and angular size in
equation~(\ref{spheredelta}) cause large errors in the calculated Doppler
factor ($\delta \propto \nu_{t}^{1.3}$ and $\xi_{d}^{1.6}$ for $\alpha=-0.5$).
Because our measurements of these quantities are only accurate to about 25\%, we estimate
our calculated Doppler factor lower limits in Tables~\ref{spherecalc} and \ref{spheretot}
to be correct only to within a factor of two.
The Doppler factor values listed in Table~\ref{spheretot} confirm this:
the average Doppler factor is 23, and the measured values show about a factor
of two scatter around this value, with most falling between 10 and 40.
There is some indication that the Doppler factor increases with time from 1991 to 1997,
but given the errors in the calculated Doppler factor, this may not be
significant.  The calculated relative X-ray brightness of components in 
turn depends on the Doppler factor from Table~\ref{spheretot}
to a high power ($S_{x} \propto \delta^{5}$ for $\alpha=-0.5$),
so the relative X-ray brightnesses of components can be in error by as much as 
a factor of 30.  This means that any of the components listed as producing more than
several percent of the X-ray emission at a given epoch in Table~\ref{spheretot} is a candidate for
producing most of the X-rays.  At eight of the epochs listed in Table~\ref{spheretot}, the
calculations show that the core produces at least two orders of magnitude
more X-rays than the next brightest component, and the calculations show the core to be the leading
X-ray producer at 15 of the 18 epochs.  Despite the large errors in the calculations, we can conclude that
in the homogeneous sphere model,
the VLBI core is the dominant source of the X-ray emission from late 1992 through 1997.
It appears that C4 dominated the X-ray emission for a time prior to 1992
(due to its small size at that time),
and that there may also be contributions at about the 10\% level from
whatever inner-jet component has just emerged from the core (this changes from C6 to
C9 over the course of the observations).

\subsection{Inhomogeneous Conical Jet Model for Core}
\label{inhomjet}

\begin{table*}[!t]
\caption{Fitted values of Conical-Jet Spectral Parameters}
\label{amn}
\begin{center}
\begin{tabular}{l c c c c c c c c} \tableline \tableline
Epoch & $\alpha$ & $m$ & $n$ & $\alpha_{s1}$ & $\alpha_{s2}$ & $\alpha_{s3}$ & $\alpha_{c2}$ & $k_{m}$ \\ \tableline \\
P1  & -0.4 & 1.6 & 1.5 & 0.21 & -0.66 & -1.31 & -0.57 & 1.2 \\
P2  & -0.6 & 2.0 & 0.8 & 0.18 & -0.85 & -1.60 & -0.68 & 1.3 \\
P3b & -0.3 & 1.8 & 1.5 & 0.33 & -0.55 & -1.28 & -0.46 & 1.3 \\
P4  & -0.4 & 1.7 & 1.4 & 0.22 & -0.65 & -1.35 & -0.55 & 1.3 \\
P5a & -0.5 & 2.0 & 1.1 & 0.30 & -0.78 & -1.45 & -0.62 & 1.4 \\
P6a & -0.5 & 1.9 & 1.3 & 0.34 & -0.81 & -1.37 & -0.66 & 1.4 \\
P6b & -0.5 & 1.9 & 1.4 & 0.39 & -0.84 & -1.32 & -0.68 & 1.4 \\ \tableline
\end{tabular}
\end{center}
\end{table*}

In this subsection we consider an alternate geometry for the VLBI core,
the inhomogeneous jet model of K\"{o}nigl (1981).  The jet
is represented by a cone with opening half-angle $\phi$, and the axis of the
jet makes an angle $\theta$ with the line-of-sight ($\theta>\phi$).
The bulk Lorentz factor of the jet is $\gamma$, the electron Lorentz factor
is $\gamma_{e}$.  Electron Lorentz factors lie between the limits $\gamma_{el}$ and $\gamma_{eu}$,
where $\gamma_{el}$ is set to 100 for all models.
The magnetic field and electron number density (or electrons plus positrons) are determined by
$B=B_{1}r^{-m}$ and $n_{e}(\gamma_{e})=K_{e}r^{-n}\gamma_{e}^{(2\alpha_{e}-1)}$,
where $r$ is the distance in parsecs from the apex of the jet,
$\alpha_{e}=\alpha$ and $K_{e}=K1$ for $\gamma_{e}<\gamma_{eb}$, and
$\alpha_{e}=\alpha-0.5$ and $K_{e}=K1\gamma_{eb}$ for $\gamma_{e}>\gamma_{eb}$,
where $\gamma_{eb}$ is a function of $r$ ($\gamma_{el}<\gamma_{eb}<\gamma_{eu}$).
The location of the break in the power-law is estimated by equating the jet travel time to a distance
$r$ with the synchrotron cooling time; see equation (4) of K\"{o}nigl (1981) and
equation (21) of Blandford \& K\"{o}nigl (1979).
The VLBI `core' emission is then due to the integrated spectrum of
the unresolved conical jet, and the position and size of the core change with
frequency in a predictable way.  
At low frequencies, the dominant emission region follows the local turnover frequency
and moves in with increasing frequency ($r\propto\nu^{-1/k_{m}}$, where
$k_{m}=((3-2\alpha)m+2n-2)/(5-2\alpha)$), and the spectral index of the integrated spectrum
is $\alpha_{s1}$.  At radius $r_{M}$ and frequency $\nu_{t}$, the local break frequency becomes
less than the turnover frequency, and the dominant emission region begins to follow the local
break frequency and moves out with increasing frequency ($r\propto\nu^{1/k_{b}}$, where
$k_{b}=3m-2$), until $\gamma_{eb}=\gamma_{eu}$ at frequency $\nu_{b}$ 
and radius $r=(\nu_{b}/\nu_{t})^{1/k_{b}}r_{M}$, which
we set to be $r_{u}$, the upper radius of the K\"{o}nigl jet.  We use this set of conditions to determine
$\gamma_{eu}$ in our models.  The spectral index of the integrated spectrum in this region is $\alpha_{s2}$.
Above $\nu_{b}$, the dominant emission region moves in with frequency, following the increasing
magnetic field ($r\propto\nu^{-1/m}$), and the integrated synchrotron spectrum falls with index $\alpha_{s3}$.
We extend the calculated emission from the upper synchrotron 
branch interior to $r_{M}$, and that from the lower synchrotron branch exterior to 
$r_{u}$, rather than truncating the spectrum at these radii as K\"{o}nigl does.
The reader is referred to Hutter \& Mufson (1986) for an illuminating diagram of this model
in the radius vs. frequency plane.

\subsubsection{Calculation of $\alpha$, $m$, and $n$}
The parameters $\alpha$, $m$, and $n$ completely determine the synchrotron and
SSC spectral indices of the integrated spectrum by the following equations (K\"{o}nigl 1981):
\begin{equation}
\alpha_{s1}=\frac{-15+4\alpha+5m-4\alpha m+5n}{3m-2\alpha m+2n-2}
\end{equation}

\begin{equation}
\alpha_{s2}=\rm{max} \left[ \begin{array}{c}
\frac{4\alpha m-2\alpha-m-n+3}{3m-2} \\ \alpha-0.5
\end{array} \right]
\end{equation}

\begin{equation}
\alpha_{s3}=-(m+2-n)/m
\end{equation}

\begin{equation}
\alpha_{c1}=\rm{min} \left[ \begin{array}{c}
\frac{-4\alpha m-2\alpha n+6\alpha+5m+10n-20}{3m-2\alpha m+2n-2} \\ 1
\end{array} \right]
\end{equation}

\begin{equation}
\alpha_{c2}=\alpha-\frac{(1-\alpha)m+2n-4}{7m-4}
\end{equation}

The SSC index $\alpha_{c1}$ (associated with the optically thick portion of the synchrotron spectrum)
is not observable in the multiwavelength spectrum because
the SSC emission in this frequency range lies well below the synchrotron emission.
The SSC index $\alpha_{c2}$ (associated with the optically thin portion of the synchrotron spectrum)
is the SSC index that is actually observed in the X-ray portion of the multiwavelength spectrum.

Since the K\"{o}nigl model parameters are so closely
tied to the multiwavelength spectrum, we make one calculation of $\delta$ for each multiwavelength
epoch listed in Table~\ref{mwspectra}, rather than one for each VLBI epoch as in $\S$~\ref{sphere}.
We fit values of $\alpha$, $m$, and $n$ at each epoch by finding the combination that most
closely reproduces the four observable spectral indices, given in Table~\ref{mwspectra} for
each multiwavelength epoch.  We take into account the errors on the observed spectral indices
given in Table~\ref{mwspectra}, and find the values of $\alpha$, $m$, and $n$ that minimize
$\chi^{2}$ at each epoch.  

Fitted values of $\alpha$, $m$, and $n$ are given in Table~\ref{amn},
along with the spectral indices calculated from these best-fit values, for comparison with Table~\ref{mwspectra}. 
In some cases, only modest agreement could be made with some of the observed indices. 
Note that the observed $\alpha_{s1}$ is an integrated value for the source that includes some steep-spectrum
jet emission.  The observed $\alpha_{s1}$ is therefore a only limiting value for the core $\alpha_{s1}$, and later we modify
the fits so the radio emission from the K\"{o}nigl model lies slightly below the total observed radio emission.
The magnetic field index $m$ is restricted to lie between 1 (for a purely transverse field) and 2
(for a purely longitudinal field).
Fitted values for $\alpha$ lie between $-0.3$ and $-0.6$, those for $m$ between 1.6 and 2.0,
and those for $n$ between 0.8 and 1.5.  In Table~\ref{amn} we also give K\"{o}nigl's parameter $k_{m}$
that controls the frequency dependent size of the core, which is proportional
to $\nu^{-1/k_{m}}$. This provides an independent check on the model fits, and using the mean core sizes
at 22 and 43 GHz from Paper I we calculate a $k_{m}$ of 1.2, in good agreement with the mean
$k_{m}$ of 1.3 from Table~\ref{amn}.

\subsubsection{Calculation of Jet Doppler Factor}
The fitted values of $\alpha$, $m$, and $n$ from Table~\ref{amn},
along with other observable parameters, were used as inputs to a K\"{o}nigl
model calculation at each epoch, implemented in the Mathcad software package.  
The other input parameters were:
the synchrotron turnover frequency and flux density and the synchrotron break frequency
from Table~\ref{mwspectra}, the observed superluminal speed, and the projected
distance of the VLBI core from the apex of the jet, $r_{proj}$.
The angle of the jet to the line-of-sight
was then varied until the predicted X-ray flux density from the K\"{o}nigl jet,
including both synchrotron and SSC flux density,  matched the
observed X-ray flux density $S_{x}$ at frequency $\nu_{x}$ from Table~\ref{mwspectra}.
Once the angle to the line-of-sight is known, the observed superluminal speed then determines
the intrinsic jet speed and Doppler factor, and the opening half-angle of the jet $\phi$ can
be calculated from equations in K\"{o}nigl (1981).

\begin{table*}[!t]
\caption{Results from Conical-Jet Model Fits}
\label{konout}
\begin{center}
\begin{tabular}{l c c c c c c c c c c c c} \tableline \tableline
& $r_{proj}$ & $\phi_{app}$ & $\xi_{r}$ & $\theta$ & & & $r_{M}$ & $r_{u}$
& $B$\tablenotemark{\dagger} & $n_{e}$\tablenotemark{\dagger} \\
Epoch\tablenotemark{\ast} & (mas) & (deg) & (mas) & (deg) & $\gamma$
& $\delta$ & (pc) & (pc) & (G) & (cm$^{-3}$) & $U_{e}/U_{B}(r_{M})$\tablenotemark{\ddagger}
& $U_{e}/U_{B}(r_{u})$\tablenotemark{\ddagger} \\ \tableline \\
P1   & 0.27 & 11.6 & 0.055 & 14.0 & 5.5 & 4.0  & 2.6  & 15.8 & 0.23 & 42  & 5.1 & 338 \\
P1*  & 0.18 & 14.4 & 0.045 & 11.8 & 5.3 & 4.9  & 1.5  & 7.7  & 0.30 & 73  & 4.8 & 234 \\
P2   & 0.28 & 11.8 & 0.058 & 6.5  & 5.8 & 8.1  & 6.3  & 25.7 & 0.17 & 5.5 & 0.9 & 168 \\
P2*  & 0.19 & 13.6 & 0.045 & 4.9  & 6.3 & 9.8  & 4.1  & 15.2 & 0.22 & 9.7 & 0.9 & 122 \\
P3b  & 0.21 & 13.2 & 0.048 & 3.3  & 7.3 & 12.3 & 10.1 & 28.8 & 0.17 & 1.3 & 0.3 & 6   \\
P3b* & 0.16 & 11.7 & 0.033 & 2.6  & 8.1 & 14.3 & 7.4  & 18.6 & 0.21 & 3.3 & 0.4 & 7   \\
P4   & 0.31 & 10.9 & 0.059 & 10.0 & 5.3 & 5.7  & 6.2  & 29.1 & 0.16 & 4.2 & 0.8 & 142 \\
P4*  & 0.22 & 11.4 & 0.044 & 7.8  & 5.5 & 7.0  & 4.0  & 16.7 & 0.21 & 8.5 & 1.1 & 51  \\
P5a  & 0.36 & 9.4  & 0.059 & 7.8  & 5.5 & 7.1  & 7.8  & 29.0 & 0.14 & 7.6 & 2.0 & 223 \\
P5a* & 0.23 & 12.9 & 0.051 & 6.3  & 5.8 & 8.3  & 4.5  & 15.1 & 0.19 & 12  & 1.7 & 126 \\
P6a  & 0.50 & 7.6  & 0.066 & 10.0 & 5.3 & 5.7  & 10.4 & 49.4 & 0.12 & 6.8 & 2.7 & 343 \\
P6a* & 0.31 & 10.5 & 0.057 & 8.2  & 5.5 & 6.8  & 5.8  & 24.8 & 0.16 & 11  & 2.2 & 205 \\
P6b  & 0.45 & 8.4  & 0.066 & 5.1  & 6.2 & 9.5  & 18.4 & 143  & 0.10 & 1.5 & 0.7 & 323 \\
P6b* & 0.28 & 10.9 & 0.053 & 3.5  & 7.2 & 12.1 & 12.6 & 87.2 & 0.14 & 2.0 & 0.5 & 173 \\ \tableline
\end{tabular}
\end{center}
\tablenotetext{\ast}{Models indicated by an asterisk have $\nu_{t}$ increased by a factor of 1.5
and $S_{t}$ decreased by a factor of 0.75 relative to their values in Table~\ref{mwspectra}.
This allows for some emission from the extended jet, which
causes the core radio flux density to lie slightly below the observed single-dish
radio flux density.}
\tablenotetext{\dagger}{Magnetic field and total electron number density are evaluated at radius $r_{M}$.}
\tablenotetext{\ddagger}{Ratio of electron
to magnetic energy density at $r_{M}$ and $r_{u}$.}
\end{table*}

Input parameters not taken from Table~\ref{mwspectra} were determined as follows.
The superluminal speed was taken to be the measured speed of the inner-jet components
C6, C7, C7a, and  C8 (the components closest to the core during the relevant epochs) from
Paper I.  Measured speeds of these four components are consistent with a constant average inner-jet
speed of 5.2$c$ (see Table 4 of Paper I), which was the apparent speed adopted for these model fits.
The projected distance of the VLBI core from the apex of the jet, $r_{proj}$, is not a directly
observable quantity, but does effect two observable quantities: the measured core radius
$\xi_{r}\approx r_{proj}\sin\phi/\sin\theta$, and the apparent jet opening angle 
$\phi_{app}\approx\phi/\sin\theta$.  In practice, $r_{proj}$ was varied until the best
fit was obtained to the observed values of these two quantities:
$\xi_{r}=0.059\pm0.018$ mas (using the mean FWHM of the Gaussian model fits from
Paper I, and multiplying by 1.6 to convert to the diameter of an optically thick sphere [Pearson 1995]),
and $\phi_{app}=10.5\pm5.3\arcdeg$ (using the mean opening angle given by component C4 from Paper I,
with C4's Gaussian FWHM converted to the diameter of an optically thin sphere for consistency). 

\begin{figure*}[!t]
\plotfiddle{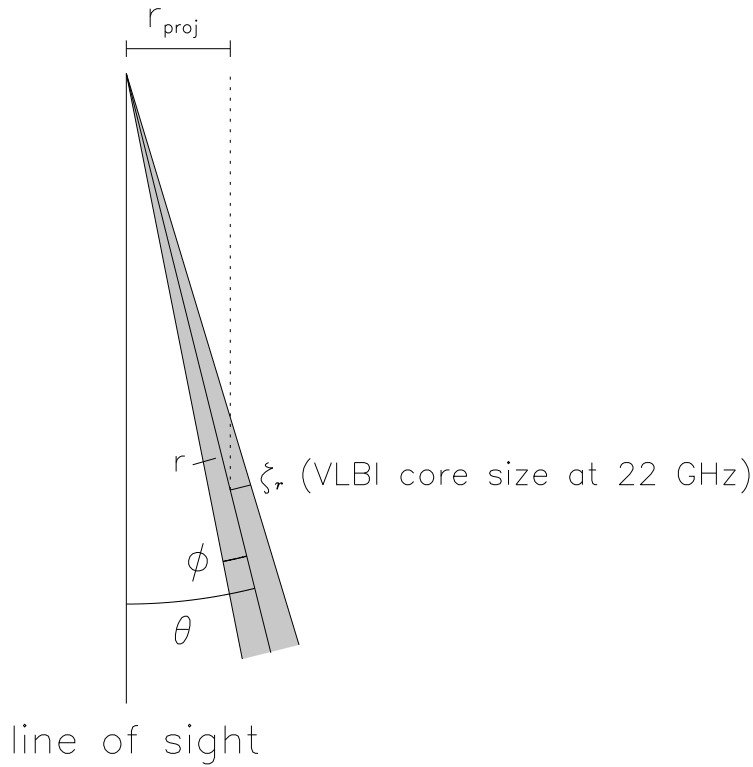}{4.4in}{0}{100}{100}{-300}{-375}
\caption{Geometry of the conical-jet model, with numerical values taken from the first
row of Table~\ref{konout} (the fit to the 1991 Jun 15 --- 1991 Jun 28 (P1) epoch).
Note that $\phi$ is the opening half-angle of the jet, $\xi_{r}$ is the jet
cross-sectional radius, $r$ is the linear distance along the jet, and $r_{proj}$
is the projected linear distance along the jet.}
\end{figure*}

Results from this model fitting, including the adopted value of $r_{proj}$ and the calculated
values of $\phi_{app}$, $\xi_{r}$, $\theta$, $\delta$, and
$\gamma$, are given in Table~\ref{konout}.
Also given in Table~\ref{konout} are $r_{M}$ (the smallest radius from which synchrotron emission
with index $\alpha$ is observed), $r_{u}$ (in our implementation, the largest radius from which
synchrotron emission with index $\alpha-0.5$ is observed),
the magnetic field and relativistic-particle number density at $r_{M}$, and the ratio of relativistic particle
to magnetic energy density at $r_{M}$ and $r_{u}$.
A sketch of a sample geometry from Table~\ref{konout} is shown in Figure 5.
The model synchrotron and SSC spectra are shown in Figure 6, where the sums of
these spectra are compared with the observed multiwavelength spectra.

In reality, some of the radio emission from 3C~279 is due to the parsec-scale jet that is resolved
by VLBI observations, and not to the partially resolved core.  To take this into account, we
constructed an alternate set of model fits with $\nu_{t}$ increased by a factor of 1.5 
and $S_{t}$ decreased by a factor of 0.75 (relative to their values in Table~\ref{mwspectra}).
This causes the predicted core radio flux density to lie slightly below the observed single-dish
radio flux densities, allowing for some emission from the extended jet  
(Piner et al. 2000; de Pater \& Perley 1983), which does not contribute significantly
to the integrated spectrum above the turnover frequency.  
In these models the jet contributes $\sim$ 3 Jy at 5 GHz,
and these model fits are
indicated by an asterisk next to the epoch name in Table~\ref{konout}.
This set of models is used in all subsequent calculations.

\begin{figure*}
\plotfiddle{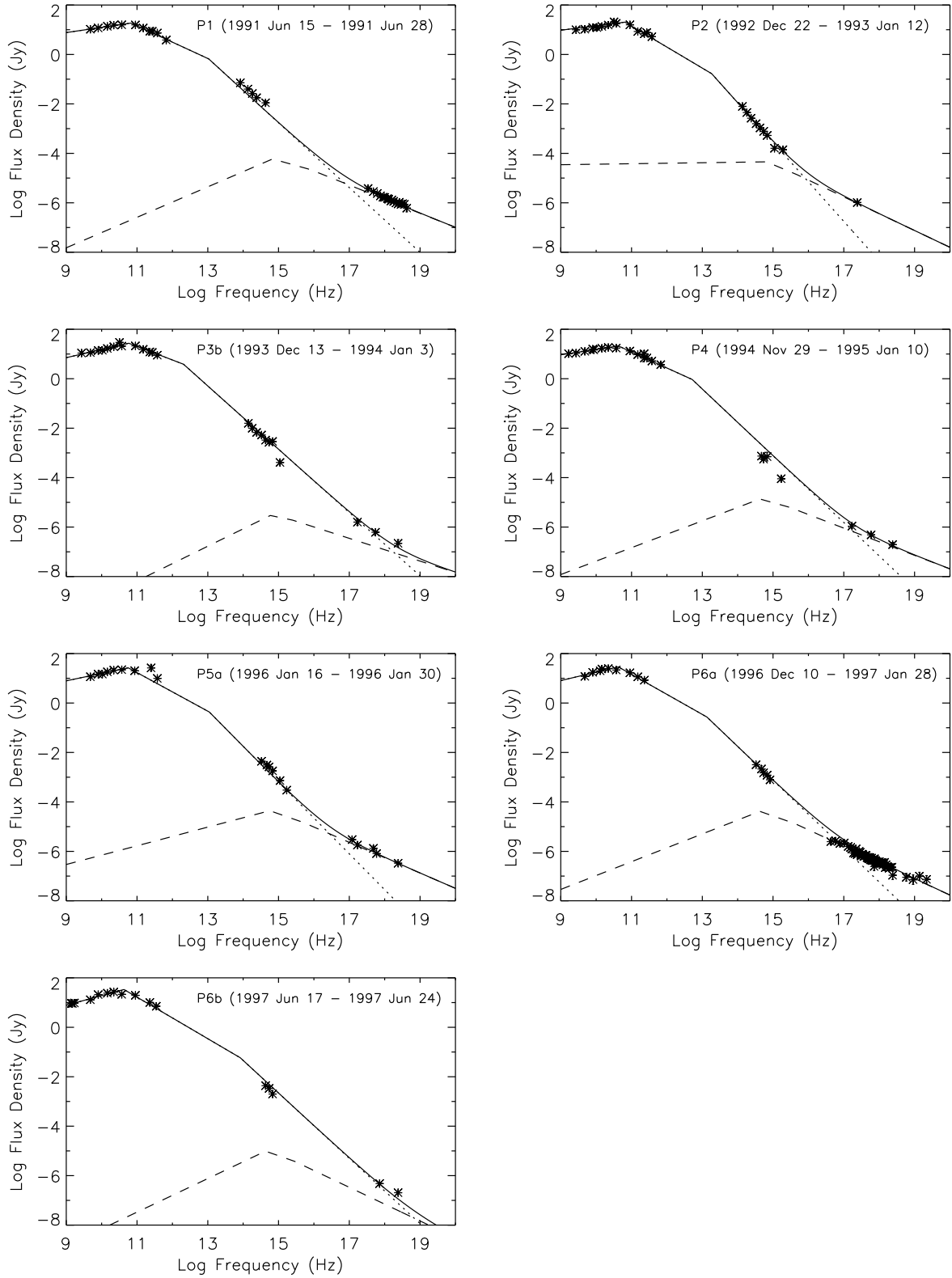}{8.0in}{0}{80}{80}{-250}{0}
\caption{Sum of the calculated synchrotron and inverse-Compton spectra 
from the K\"{o}nigl model fit to each multiwavelength epoch, compared
with the observed flux densities at each multiwavelength epoch, indicated by asterisks.
The spectral sum is indicated by the solid line, the synchrotron spectrum by the dotted line,
and the inverse-Compton spectrum by the dashed line.}
\end{figure*}

The agreement between the calculated and observed spectra is reasonably good, although there
are discrepancies (particularly with the optical flux densities, e.g., 1994 Nov 29 --- 1995 Jan 10,
period P4 in H01) of about a factor of 2.
There are several reasons for this: the optical flux density of 3C~279 is known to vary by
as much as 60\% on timescales of a day (Balonek \& Kartaltepe 2002)
so variability may be a factor, in some cases one of the
four observed spectral indices was poorly matched by the three free parameters $\alpha$, $m$, and $n$, and
at some epochs there are apparently significant synchrotron X-rays, so the observed $\alpha_{c2}$ is
not an accurate measurement of the Compton spectral index at those epochs.
The spectra in Figure 6 do not extend up to the EGRET energy range, and we do not attempt
to model the high-energy $\gamma$-ray emission with the K\"{o}nigl model.  
Although the K\"{o}nigl model can produce $\gamma$-rays in the EGRET energy range,
there is considerable evidence that the GeV emission is external Compton and not SSC
(Kubo et al. 1998; H01; Ballo et al. 2002),
so we do not apply the EGRET data as a constraint on our SSC model.

The values of the Doppler factor obtained by applying a conical geometry to the core are
about a factor of two lower than the Doppler factors obtained using a spherical geometry ($\S$~\ref{sphere}),
showing that the assumed geometry can have a moderate influence on the SSC Doppler factor.
The average Doppler factor from the fits in Table~\ref{konout} that allow for emission from
an extended jet is 9 with a scatter of $\pm3$.
When inputs to the model are all varied by 10\% (such that the changes all act together
to increase or decrease the Doppler factor), the fitted $\delta$ can be made to vary
by about 30\%.  
Thus variations in the Doppler factor from epoch to epoch in Table~\ref{konout} could be due to the
observational uncertainties in the input parameters, rather than real variations
in the Doppler factor.

\subsection{Combined Conical-Jet/Sphere Models}
\label{combined}
We are now in a position to make a model for the source that combines the conical VLBI core
with the homogeneous sphere VLBI jet components ($\S$~\ref{sphere}).  We use the homogeneous
sphere values from Table~\ref{spherecalc}, and the conical-jet core values from Table~\ref{konout},
using the multiwavelength epoch closest in time to each VLBI epoch.  As in $\S$~\ref{sphere}, each
calculated Doppler factor (either conical jet or sphere) provides a lower limit to the Doppler
factor at that epoch (assuming a constant Doppler factor along the jet),
since each calculation assumes the component under consideration produces
all of the X-rays.  We solve for {\em one} Doppler factor at each epoch as follows.
We lower the X-ray flux density attributed to the conical-jet core in the K\"{o}nigl-jet fitting program,
while adjusting $r_{proj}$ to maintain the observed values of $\phi_{app}$ and $\xi_{r}$.
This has the effect of raising the calculated Doppler factor.  We use this Doppler factor
to calculate the X-ray flux density that would be observed from the jet components in Table~\ref{spherecalc},
and add these flux densities to that attributed to the VLBI core.  We continue this process until we
find a total X-ray flux density that equals the observed X-ray flux density.
Results of this process are given in Table~\ref{jettot}.
The parameter space of the K\"{o}nigl jet model is complicated, and there is not a simple relation
between the observed X-ray flux density and $\delta$ as for the homogeneous sphere.
In most regions of parameter space $\delta$ is a slowly varying function of the X-ray flux density as
for a sphere, but there are regions where $\delta$ can change dramatically in response to a small
change in the X-ray flux density (e.g., the last two entries in Table~\ref{jettot}, where, as the
solution for $\theta$ begins to approach zero, the Lorentz factor increases to maintain the observed
superluminal speed, producing the high Doppler factors in these models).

\begin{table*}[!t]
\caption{X-Ray Production in the Combined Conical-Jet/Sphere Model}
\label{jettot}
\begin{center}
\begin{tabular}{l r r r r r r r r r c c} \tableline \tableline
& & \multicolumn{8}{c}{Fraction of Total X-Rays by Component} & $\theta^{\ddagger}$ \\
Epoch & $\delta$ & Core & C4   & C6   & C6/7$^{\dagger}$ & C7
& C7a/C8$^{\dagger}$ & C8   & C8/9$^{\dagger}$
& (deg) & $\gamma^{\ddagger}$ \\ \tableline \\
1991 Jun 24 & 16 & 0.09 & 0.91 & ...  & ...  & ...  & ...  & ...  & ...  & 2.8 &  9.8 \\
1992 Jun 14 & 10 & 0.83 & 0.17 & ...  & ...  & ...  & ...  & ...  & ...  & 4.4 &  6.6 \\
1992 Nov 10 & 10 & 0.97 & 0.02 & 0.01 & ...  & ...  & ...  & ...  & ...  & 4.8 &  6.4 \\
1993 Feb 17 & 10 & 0.99 & 0.01 & ...  & ...  & ...  & ...  & ...  & ...  & 4.9 &  6.3 \\
1993 Nov 8  & 14 & 1.00 & ...  & ...  & ...  & ...  & ...  & ...  & ...  & 2.6 &  8.1 \\
1994 Mar 2  & 15 & 0.95 & ...  & ...  & ...  & 0.05 & ...  & ...  & ...  & 2.3 &  8.6 \\
1994 Jun 12 & 8  & 0.77 & 0.15 & ...  & ...  & 0.08 & ...  & ...  & ...  & 6.6 &  5.7 \\
1994 Sep 21 & 8  & 0.84 & 0.01 & ...  & 0.15 & ...  & ...  & ...  & ...  & 6.9 &  5.7 \\
1995 Jan 4  & 8  & 0.72 & 0.07 & 0.05 & ...  & 0.16 & ...  & ...  & ...  & 6.3 &  5.8 \\
1995 Feb 25 & 13 & 0.36 & 0.01 & ...  & ...  & 0.63 & ...  & ...  & ...  & 3.2 &  7.4 \\
1995 Mar 19 & 9  & 0.57 & 0.06 & ...  & ...  & 0.37 & ...  & ...  & ...  & 5.3 &  6.1 \\
1996 Jan 7  & 11 & 0.48 & 0.51 & ...  & ...  & ...  & 0.01 & ...  & ...  & 4.9 &  8.1 \\
1996 May 13 & 12 & 0.28 & 0.16 & ...  & ...  & ...  & 0.56 & ...  & ...  & 3.4 &  7.2 \\
1996 Jun 9  & 19 & 0.08 & 0.18 & ...  & ...  & ...  & 0.74 & ...  & ...  & 1.6 & 10.2 \\
1997 Jan 15 & 26 & 0.08 & 0.77 & ...  & ...  & ...  & ...  & 0.15 & ...  & 1.2 & 14.1 \\
1997 Mar 29 & 37 & 0.09 & 0.72 & ...  & ...  & ...  & ...  & ...  & 0.19 & 0.6 & 19.3 \\
1997 Jul 16 & 27 & 0.87 & 0.11 & ...  & ...  & ...  & ...  & ...  & 0.02 & 0.8 & 13.8 \\
1997 Nov 16 & 33 & 0.85 & 0.15 & ...  & ...  & ...  & ...  & ...  & ...  & 0.5 & 16.8 \\ [5pt] \tableline
\end{tabular}
\end{center}
\hspace{0.50in} $\dagger$ Refers to a blended component.\\
\hspace*{0.50in} $\ddagger$ These values are discussed in $\S$~\ref{speedor}.
\end{table*}

We reiterate that the predicted X-ray emission depends strongly on the observed quantities,
so that any of the components listed as producing more than
several percent of the X-ray emission at a given epoch in Table~\ref{jettot} is a candidate for
producing most of the X-rays.  Because of this, it is better to consider averages over
many epochs than the entry in Table~\ref{jettot} from a single epoch.  On average, modeling the
core as a conical jet rather than a homogeneous sphere reduces its contribution to the X-ray flux density.
When the core is modeled as a conical jet, then on average the core produces about half of the
X-rays, with the other half being split about evenly between C4 and the brightest inner-jet component.

\section{Discussion}
\subsection{Comparison to Other Estimates of $\delta$ for 3C~279}
\label{compdelta}
A strong lower limit to the Doppler factor can be obtained by enforcing the condition
that the emitting region should be transparent to $\gamma$-rays.  Inferring a source
size from the $\gamma$-ray variability timescale during the large flare in early 1996,
Wehrle et al. (1998) find $\delta>6.3$ for 1 GeV photons and $\delta>8.5$ for 10 GeV photons.
This agrees very well with the lower limit to the core $\delta$ found at this epoch in this
paper, $\delta>8.3$ (Table~\ref{konout}), and with the overall $\delta$ found
at this epoch, $\delta$=11 (Table~\ref{jettot}).

An independent method for measuring $\delta$ is to compare the radio core brightness temperatures
measured from VLBI maps and from radio light curves (L\"{a}hteenm\"{a}ki et al. 1999).
Since these depend on the intrinsic brightness temperature multiplied by $\delta$ raised
to different powers, a measured value of $\delta$ can be extracted from these two observables.
We applied this method to 3C~279 in Paper I and found $\delta$=7.4 at epoch 1995.2.
Again, this agrees very well with the lower limit to the core $\delta$ found at this epoch in this
paper, $\delta>7.0$ (Table~\ref{konout}), and with the overall $\delta$ found
at this epoch, $\delta$=9 (Table~\ref{jettot}).  The similar Doppler factors found from these
two independent methods differ from the much higher Doppler factor ($\delta\sim 100$) found when equipartition
between magnetic and particle energy is assumed (Paper I).  Other calculations of an equipartition
Doppler factor for 3C~279 (e.g. G\"{u}ijosa \& Daly 1996) have found a lower 
equipartition Doppler factor because they did not have access to the high-resolution 
(and high brightness temperature sensitivity) VLBI data presented in Paper I.

Models using various emission processes and geometries
to explain the multiwavelength spectra of 3C~279, including the $\gamma$-ray emission,
have constrained $\delta$ as part of their model fitting.  
Some examples are as follows:  Maraschi, Ghisellini, \& Celotti (1992)
used an accelerating parabolic jet where $\delta$ varied from 10 to 18.
Ghisellini \& Madau (1996) assumed $\delta$=14 for application of their ``mirror''
model to the high-energy emission of 3C~279.  H01 and Ballo et al. (2002) find Doppler factors 
ranging from 8 to 23 and from 12 to 19 respectively for application of their models to the various 
multiwavelength spectra presented in those papers.
In these cases the angle to the line-of-sight and/or the Lorentz factor are assumed input
quantities, so the derived Doppler factors are merely consistent within the framework
of the particular model, not actual Doppler factor measurements.
In addition, these models do not apply constraints from the VLBI observations
as we do in this paper, so they predict features in the VLBI maps that are not observed.
The models mentioned in this paragraph predict apparent speeds in the inner jet ranging
from 1 to 24$c$, whereas the observations show apparent speeds in the inner jet to be about 5$c$ (Paper I),
which is enforced by all of the models in Table~\ref{konout}.

Note that our reliance on the apparent superluminal speed assumes that the pattern
speed observed in the VLBI observations (Paper I) is equal to the bulk fluid speed.
While there is one stationary component (C5) that clearly does not move at the bulk fluid speed,
all five components observed in the inner jet of 3C~279 (C5a, C6, C7, C7a, and C8)
during the course of the monitoring described in Paper I moved with approximately the
same apparent speed of 5$c$, so we take this value as an indicator of the apparent fluid speed
in the inner jet.  If the apparent speed is not used as an observable, then
the speed and orientation of the jet are not tightly constrained in our models.  For example, allowing the
apparent bulk speed to range from 1 to 10$c$ allows the following ranges of parameters for the
P1 model in Table~\ref{konout}: $5\arcdeg<\theta<14\arcdeg$, $3<\gamma<24$, and $2<\delta<4$.

\subsection{Implications of K\"{o}nigl Jet Model}

\subsubsection{Comparison to Sphere Model Results}
In comparing conical and spherical models for the VLBI core, we find
that a smaller Doppler factor is required to reduce the predicted core X-ray flux
density to the measured X-ray flux density using the conical geometry.
This implies that, if a conical geometry is indeed the correct geometry
for the VLBI core, inverse-Compton calculations such as those of Ghisellini et al. (1993)
that have assumed a spherical geometry may have systematically
overestimated $\delta$ in their samples.

\subsubsection{Implications of $\alpha$, $m$, and $n$}
The values of $\alpha$, $m$, and $n$ derived for the K\"{o}nigl model
determine the orientation of the magnetic field and the ratio of
relativistic particle to magnetic energy density.  The value of $m$ can vary between
1 and 2 in the K\"{o}nigl model, with $m=1$ corresponding to a purely transverse
magnetic field, and $m=2$ to a purely longitudinal field.  
Conservation of particle number in a conical jet requires that $n=2$,
our values of $n<2$ imply that the total number of relativistic particles increases down the jet
(e.g., from continuous acceleration adding to the total number of relativistic electrons).
Our fitted values of $\alpha$ vary between $-0.3$ and $-0.6$,
our values of $n$ between 0.8 and 1.5, and our values of $m$ between 1.6 and 2.0, 
which corresponds to a predominantly longitudinal magnetic field in the region of
the jet modeled by the K\"{o}nigl model, which is about 0.1 mas in size for 3C~279. 
In contrast, VLBI polarimetry observations at many frequencies from 15 to 86 GHz (Lepp\"{a}nen, Zensus, \& Diamond 1995;
Taylor 1998; Lister, Marscher, \& Gear 1998; Homan \& Wardle 1999; Lister \& Smith 2000; Attridge 2001)
have given magnetic field vectors oriented perpendicular to the jet in the core region.
We do not consider these results to be in conflict, since the VLBI
polarimetry observations show the core to have low levels of polarization even at
86 GHz (Attridge 2001), suggestive of an initially tangled field that becomes ordered at shocks outside
the K\"{o}nigl jet region.  VLBI polarimetry with beams smaller than 0.1 mas would
be needed to image the magnetic field structure in the region modeled in this paper.

The values of $m$ and $n$ also determine how the ratio of
relativistic particle to magnetic energy density varies along the jet (it goes approximately as $r^{-n+2m}$),
this ratio is quoted in Table~\ref{konout}
at radii $r_{M}$ and $r_{u}$.  The jet is close to equipartition at $r_{M}$,
but the degree of particle dominance increases down the jet, and at $r_{u}$
the relativistic-particle energy density dominates the magnetic energy density
by roughly two orders of magnitude.
The increasing particle dominance of K\"{o}nigl jets with jet radius seems to be a common feature
of these models when $n$ and $m$ are determined from spectral fits (rather than assumed).  Similar
particle dominance was found by Unwin et al. (1994, 1997) and by Hutter and Mufson (1986), who
attributed the result qualitatively to a conversion of magnetic energy to particle energy by
magnetohydrodynamic jet acceleration.
If the K\"{o}nigl model is a
correct description of the VLBI core of 3C~279, then this core is not in equipartition,
which could explain why 3C~279's equipartition Doppler factor is so much higher than the
Doppler factor measurements obtained by other means (see $\S$~\ref{compdelta}).
This finding contrasts with that of L\"{a}hteenm\"{a}ki et al. (1999), who find that
Doppler factors computed by comparing variability and VLBI brightness temperatures
in general agree with equipartition Doppler factors, although analysis of brightness
temperature measurements from the high-resolution VLBI data of Paper I has shown that this is not the case for 3C~279.

From the particle energy density, we can calculate the energy flux associated
with the particles from
\begin{equation}
\label{kinflux}
L_{kin}\approx \frac{4}{3}\pi c r_{jet}^{2} \gamma^{2} U_{e} (1+k),
\end{equation}
where $r_{jet}$ is the linear size of the jet cross-sectional radius,
$U_{e}$ is the relativistic-particle energy density (see Table~\ref{konout}), and 
$k$ is the ratio of proton to electron energy.
See De Young (2002) equation (4.107), Bicknell (1994) equation (52),
and Celotti \& Fabian (1993) equation (1) for discussions of this equation.
Note that $U_{e}$ depends on $\gamma_{el}$, which is not precisely known
(although the observed lack of Faraday rotation in most extragalactic radio sources
implies $\gamma_{el}>100$, Jones \& O'Dell 1977),
so that this is only an order-of-magnitude calculation.
From equation~\ref{kinflux},
the particle energy flux is of order 10$^{46} (1+k)$ ergs s$^{-1}$.
This is of the same order as the particle kinetic energy fluxes found by Celotti \& Fabian (1993),
for a sample of sources that included 3C~279.
The particle energy flux is about an order of magnitude higher than the bolometric radiative 
luminosity of 3C~279's jet (Hartman et al. 1996) after correction for beaming,
which falls within the range of $L_{kin}/L_{rad}$ found by Celotti \& Fabian (1993).
An energy flux of 10$^{46} (1+k)$ ergs s$^{-1}$ is
equivalent to an energy injection rate 
of order $0.1 (1+k) M_{\odot}$ yr$^{-1}$, or a mass accretion rate of order $0.1 (1+k)/\eta M_{\odot}$ yr$^{-1}$,
where $\eta$ is the efficiency of conversion of mass to kinetic energy.

We can compare the values of $\alpha$, $m$, and $n$ found here for 3C~279 with values
of these parameters found for other sources where the K\"{o}nigl model has been applied.
The source apart from 3C~279 with the most constraints from spectral and VLBI data is 3C~345.
Unwin et al. (1994) found $\alpha=-0.6$, $m=1.5$, and $n=1.4$ for 3C~345 in mid-1990.
Unwin et al. (1997) found $\alpha=-0.6$, $m=1.9$, and $n=1.7$ for 3C~345 in mid-1992, and at this
epoch they found that the K\"{o}nigl-jet core was not the dominant X-ray emitter in the source.
Hutter \& Mufson (1986) found $1.1<m<1.6$ and $1.1<n<1.6$ with an assumed $\alpha$ of $-0.5$ in their application 
of the K\"{o}nigl model to three nearby BL Lac objects.

\subsubsection{Need for an additional homogeneous component}
The spectral index in the K\"{o}nigl model fit for 3C~279 that has the poorest
observational constraints is the synchrotron index below the turnover frequency, 
$\alpha_{s1}$.  This is because the extended jet
emits a significant fraction of the flux at low radio frequencies, so that what 
we get from the single-dish spectra shown in Figure 6 is the spectrum of the 
core plus jet, when what we want is the spectrum of the core alone.  The observed $\alpha_{s1}$
thus provides only a limit to the actual $\alpha_{s1}$, with high-resolution VLBI at
low frequencies being needed to accurately measure $\alpha_{s1}$.  There are indications from
VSOP observations at 1.6 and 5 GHz (Piner et al. 2000) that, at least at that epoch,
the spectral index of the VLBI core was much more inverted than the values of $\alpha_{s1}$
quoted in Table~\ref{mwspectra}.  If confirmed by further VSOP data on 3C~279 (Edwards et al. in preparation),
this would rule out a pure K\"{o}nigl jet model for the 3C~279 core at these epochs.  Because the K\"{o}nigl model
was created in part to explain the flat radio spectra of quasars, very inverted values of $\alpha_{s1}$
(less than about $-1.0$) create contradictions in the model, such as a synchrotron break that goes the
`wrong way' ($\alpha_{s3}>\alpha_{s2}$).  This situation could be rectified by adding a
homogeneous component (a newly emerging `blob') that is blended with the VLBI core on the VLBI images.

It seems likely that, in the general case, the spectrum of the VLBI core is a blend of
an inhomogeneous component like a K\"{o}nigl jet and one or more shocks moving along the jet.
Because these would all be merged on the VLBI images, it would be impossible to determine an inverse-Compton
Doppler factor in this case, because the crucial observational constraint provided by the VLBI size would be missing.
The degree to which the ``quiescent'' emission from 3C~279 can be represented by a single inhomogeneous component
will determine the reliability of the Doppler factors computed from the conical-jet geometry.

\subsubsection{Speed and Orientation of the Jet}
\label{speedor}
The Lorentz factor $\gamma$ and angle to the line-of-sight $\theta$ of the jet can be calculated if $\delta$
and the apparent speed $\beta_{app}$ are known:
\begin{equation}
\gamma=\frac{\beta_{app}^{2}+\delta^{2}+1}{2\delta}
\label{gameq}
\end{equation}
and
\begin{equation}
\theta=\arctan\frac{2\beta_{app}}{\beta_{app}^{2}+\delta^{2}-1}.
\label{theeq}
\end{equation}
In their similar work on 3C~345, Unwin et al. (1997) calculate the jet speed and
angle to the line-of-sight at several radii along the jet.  This was possible for 3C~345 because
the jet component C7 was the only good candidate for producing the X-ray emission, so the
inverse-Compton Doppler factor measured for C7 (assuming C7 produced 100\% of the X-rays) could be combined with the apparent
speed measured for C7 at different points along the jet to produce a plot of $\gamma$
and $\theta$ vs. $r$ (see Figure 4 of Unwin et al. 1997). 

The situation for 3C~279 is not so straightforward.  As discussed in $\S$~\ref{combined},
the core, inner-jet components, and C4 all probably contribute a non-negligible fraction
of the X-ray emission.  A unique solution for $\delta$ for each component can only be obtained by knowing
{\em a priori} what this fraction is.  Table~\ref{konout} gives Lorentz factors and 
angles to the line-of-sight for the conical-jet core under the assumption that the core produces all of the X-rays.
Average values of $\gamma$ and $\theta$ obtained in this fashion are $\gamma=6$ and $\theta=6\arcdeg$.
Better estimates of $\gamma$ and $\theta$ can be obtained by using instead the Doppler factor that
reproduces the observed X-ray emission when {\em all} components are considered (see Table~\ref{jettot}).
This provides a reasonable estimate for $\gamma$ and $\theta$ for the component listed as producing
the majority of the X-ray emission at that epoch.  These values of $\gamma$ and $\theta$ are listed
in the final two columns of Table~\ref{jettot}.  At epochs where the core or an inner-jet component
was the dominant X-ray producer we used the average apparent speed of the inner jet of 5.2$c$.
At the 4 epochs where C4 was the dominant X-ray producer we used the apparent speed of C4, or 7.5$c$ (Paper I).
In this fashion, we obtained average Lorentz factors and angles to the line-of-sight for the core and inner-jet region ($r<1$ mas)
of $\gamma=8$ and $\theta=4\arcdeg$, and an average Lorentz factor and angle to the line-of-sight for C4 (at $r\approx3$ mas) of
$\gamma=13$ and $\theta=2\arcdeg$.  These numbers apply only for a specific portion of C4's curved trajectory (see below),
along this portion of its trajectory C4 is faster, in intrinsic speed and apparent speed, and its path
is closer to the line-of-sight than the inner jet components.

Whether this speed and angle are characteristic of C4's location in the jet or are unique to C4 is
unknown, because all components other than C4 faded by the time they reached
1 mas from the core during the period of our monitoring (see Paper I for a discussion).
The quoted speed and angle for C4 apply for times when C4 was contributing significantly to the X-ray emission,
at the beginning and the end of the observed time range.  The Doppler factor of C4 is evidently
time-variable and increased around 1997, coincident with its brightening on the component light curves (Paper I).
In Paper I we presented a detailed analysis of the kinematics of C4, based on its curved trajectory in
the VLBI images.  For that analysis we assumed $\gamma=13$ (slightly higher than the minimum $\gamma$ required
for the maximum apparent speed along C4's curved path).  We confirm this choice of $\gamma$ in this
paper, and also confirm the other results from the kinematic analysis in Paper I: that the angle to the line-of-sight of C4
is about 2$\arcdeg$, and that C4's Doppler factor was highest at the beginning and end of the observed time
range (see Figure 8 of Paper I).

\section{Conclusions}
In this paper we have calculated, as accurately as is possible, inverse-Compton Doppler factors
for 3C~279.  These Doppler factors were then used to compute the speed and orientation of the parsec-scale jet.
Calculation of inverse-Compton Doppler factors is a notoriously inaccurate business.
Nevertheless, given the large amount of multiwavelength spectral data recently published by H01,
and the large amount of VLBI data recently published by us (Paper I),
3C~279 seems to be the best object for constraining the Doppler factor by this method.
Progress on this calculation for other sources is hindered mainly by the reliance on the turnover
frequency of the core and jet components, which is a critical parameter that is relatively poorly constrained
observationally, even for this well-observed source (see Figure 4).  Nearly simultaneous VLBI observations
at as many frequencies as possible (and as near as possible to the same resolution) are required before
attempting such a calculation.

Major conclusions from this work are:
\begin{enumerate}

\item{The VLBI morphology and multiwavelength data cannot be adequately explained by
either an inhomogeneous jet ($\S$~\ref{inhomjet}) or a homogeneous sphere
(or spheres) geometry ($\S$~\ref{sphere}) alone.  We have used a combined model ($\S$~\ref{combined}),
where we assume an inhomogeneous conical-jet geometry for the VLBI core
and a homogeneous sphere geometry for the VLBI components.  In this combined model,
we require the Doppler factor to be the same for the jet and spheres; the spheres
can therefore be regarded as approximations to dense clumps propagating along with the bulk jet.
By applying this method, we obtain an average speed and angle to the line-of-sight for the core and inner-jet region ($r<1$ mas)
of $v=0.992c$ ($\gamma=8$) and $\theta=4\arcdeg$, 
and an average speed and angle to the line-of-sight for C4 (at $r\approx3$ mas) of
$v=0.997c$ ($\gamma=13$) and $\theta=2\arcdeg$.}

\item{When the core is modeled as a conical jet, then on average the core produces about half of the
X-rays, with the other half being split about evenly between C4 and the brightest inner-jet component.
This result differs from that for 3C~345 found by Unwin et al. (1997), who could not match the
K\"{o}nigl model to 3C~345's X-ray emission, and concluded the core is not the dominant X-ray emitter in 3C~345.}

\item{The jet is particle dominated at most radii
that produce significant observed emission in the K\"{o}nigl model.  This result was also found by Unwin et al. (1994)
for 3C~345.  At the inner radius of the K\"{o}nigl jet the magnetic field is of order 0.1 G
and the relativistic-particle number density is of order 10 cm$^{-3}$.
The kinetic energy flux in the jet is of order
10$^{46} (1+k)$ ergs sec$^{-1}$, where $k$ is the ratio of proton to electron energy,
which implies a mass accretion rate of order $0.1 (1+k)/\eta M_{\odot}$ yr$^{-1}$,
where $\eta$ is the efficiency of conversion of mass to kinetic energy.}

\end{enumerate}

\acknowledgements
Part of the work described in this paper has been carried out at the Jet
Propulsion Laboratory, California Institute of Technology, under
contract with the National Aeronautics and Space Administration.
AEW acknowledges support from the NASA Long-Term Space Astrophysics Program.
BGP acknowledges helpful conversations with David Meier and helpful comments from the referee, and support
from the NASA Summer Faculty Fellowship Program and Whittier College's Newsom Endowment.

\end{document}